\documentclass{elsarticle}

\usepackage{amsmath,amssymb,amsthm,physics,mathrsfs,graphicx,hyperref}

\usepackage[numbers]{natbib}

\newcommand{\quantop}[1]{\mathcal{#1}}
\newcommand{\HH}{\quantop{H}}
\newcommand{\vect}[1]{\mathbf{#1}}
\newcommand{\tensor}[1]{\stackrel{\leftrightarrow}{#1}}
\newcommand{\set}[1]{\mathscr{#1}}

\newcommand{\rme}{\mathrm{e}}
\newcommand{\boxi}{\boldsymbol{\xi}}
\newcommand{\bosi}{\boldsymbol{\sigma}}

\begin{document}

\title{On computational capabilities of Ising machines based on
  nonlinear oscillators}

\author{Mikhail Erementchouk}%
\ead{merement@gmail.com}
 \author{Aditya Shukla} 
\ead{aditshuk@umich.edu }
 \author{Pinaki Mazumder}%
\ead{pinakimazum@gmail.com}
 \address{%
   Department of Electrical Engineering and Computer Science,
 University of Michigan, Ann Arbor, MI 48109 USA 
}%

\begin{abstract}
  Dynamical Ising machines are actively investigated from the perspective
  of finding efficient heuristics for NP-hard optimization problems.
  However, the existing data demonstrate super-polynomial scaling of the
  running time with the system size, which is incompatible with large
  NP-hard problems. We show that oscillator networks implementing the
  Kuramoto model of synchronization are capable of demonstrating polynomial
  scaling. The dynamics of these networks is related to the semidefinite
  programming relaxation of the Ising model ground state problem.
  Consequently, such networks, as we numerically demonstrate, are capable
  of producing the best possible approximation in polynomial time. To reach
  such performance, however, the reconstruction of the binary Ising state
  (rounding) must be specially addressed. We demonstrate that commonly
  implemented forced collapse to a close-to-Ising state may diminish the
  computational capabilities up to their complete invalidation. Therefore,
  consistent treatment of rounding may cardinally improve various operation
  metrics of already existing and upcoming dynamical Ising machines.
\end{abstract}


\maketitle

\section{Introduction}
\label{sec:intro}

Challenges set by large-scale NP-hard problems make unconventional
models of computation of special interest and importance. One of such
models is based on the Ising model describing a network of coupled
classical spins. In 1970--1980-s, researchers realized that reaching
the equilibrium of the Ising model is equivalent to solving certain
optimization problems \cite{kirkpatrick_optimization_1983,
  fu_application_1986, barahona_application_1988}. Furthermore, in
\cite{barahona_application_1988}, it was observed that the ground
state of the Ising model on a graph delivers the maximal cut of the
graph. This tied the Ising model with a series of other NP-hard
problems as established
in~\cite{miller_reducibility_1972,garey_simplified_1976} and
explicated in~\cite{lucas_ising_2014}. These observations exposed the
Ising model as a special model of computation, which represents
computing tasks in terms of set partitioning.

Recently, a significant research effort is put into development of
continuous dynamical Ising machines~\cite{wang_coherent_2013,
  novikov_oscillatory_2014, marandi_network_2014,
  mahboob_electromechanical_2016,inagaki_coherent_2016,
  parihar_vertex_2017, kalininNetworks2018,pierangeli_large-scale_2019,
  chou_analog_2019, mcquillan_oim_2019,wang_new_2019,
  goto_combinatorial_2019, kalinin_polaritonic_2020,
  yamamoto_coherent_2020,ahmed_probabilistic_2020, mallickUsing2020,
  duttaUnderstanding2020, afoakwa_brim_2021, goto_high-performance_2021}.
Characteristically, these machines do not represent the classical spins by
a binary (taking values $\pm1$) object. Instead, they leverage the emergent
property of specially constructed continuous dynamical system to minimize
the Ising model Hamiltonian.

To better understand challenges faced by Ising machines, it should be
reminded that finding the maximum cut is an APX-hard
problem~\cite{papadimitriou_optimization_1991, khot_optimal_2004}. If
$\mathrm{P} \ne \mathrm{NP}$, an APX-hard problem cannot be solved with
arbitrarily good approximation within time scaling polynomially with the
system size. For finding the maximal cut, this means that any Ising solver
acting on a sufficiently rich set of large graphs can either guarantee, at
best, $16/17\approx0.941$~\cite{hastad_optimal_2001} value of the maximal cut, or
its running time will scale super-polynomially. The fact that an Ising machine
is continuous does not change the implications of the APX-hardness.
Moreover, finding the ground state of the Ising model admits various
continuous (exact) representations. We consider an Ising machine based on
one of such representation and show that its \emph{performance} --- the
degree of approximation achievable in polynomial time --- is the same as of a
simple local search.

The questions of performance received little attention in the Ising machine
literature. Virtually, the only adopted method of evaluating the accuracy
is testing the machines against a selective set of benchmarks. While such
tests are, of course, important, they provide little insight into how the
machines would operate on a general class of problems, especially while
dealing with graphs of gigantic size (say, with millions or billions of
nodes)~\cite{yurtseverScalable2021}. In turn, studies of the dependence of
the running time on the problem size are also scarce. The existing
data~\cite{hamerlyExperimental2019, leleuScaling2021} demonstrate
super-polynomial scaling: the best result reported
in~\cite{leleuScaling2021} is $\mathcal{O}\left( \rme^{\sqrt{N}} \right)$, where
$N$ is the number of graph nodes. Such scaling effectively puts large
NP-hard problem out of the reach.

In the present paper, we show that Ising machines based on networks
implementing the Kuramoto model of synchronization
\cite{kuramoto_phase-_1997,mori_dissipative_1998,acebron_kuramoto_2005,
  pietras_network_2019} are capable of demonstrating scaling compatible
with large NP-hard problems. The ability of these
machines~\cite{chou_analog_2019, mcquillan_oim_2019,
  ahmed_probabilistic_2020, mallickUsing2020, duttaUnderstanding2020,
  bashar_experimental_2020, wangSolving2021} to find the ground state of
the Ising model is of the same origin as for the semidefinite programming
(SDP) relaxation \cite{gartner_approximation_2012}. This suggests that such
oscillatory Ising machines can reach the theoretical limits: the best
classically possible quality of solutions
\cite{khot_optimal_2004,raghavendraOptimal2008a} in time that scales almost
linearly with the problem size \cite{trevisanMaxCutSmallest2012,
  sotoImprovedAnalysisMaxCut2015}.

The best theoretical performance, however, is not achieved automatically
within the commonly adopted ways to implement machines based on
synchronized networks. As the dynamical model describing these machines, we
consider a network of oscillators with identical natural frequencies. In
the rotating frame, the network is governed by
\cite{shinomoto_phase_1986,wangSolving2021}
\begin{equation}\label{eq:KM-full}
 \dot{\theta}_m = K \sum_{n} A_{m,n} \sin(\theta_n - \theta_m) + K_s \sin(2 \theta_m),
\end{equation}
where $\theta_m $ is the $m$-th oscillator phase, $A_{m,n} $ is the network
adjacency matrix, $K $ is the coupling strength, and $K_s $ is the
strength of the phase injection~\cite{adler_study_1946,
  zhang_study_1992, bhansali_gen-adler_2009}, which facilitates
aligning individual oscillator phases. To distinguish from the model
with variable frequencies, we will refer to~\eqref{eq:KM-full} as the
quasistatic Kuramoto model (QKM).

The relation with the Ising model is often suggested along the
following lines. QKM can be considered as induced by the Lyapunov
function (a function monotonously decreasing with time with evolution
of the system)~\cite{shinomoto_phase_1986}
\begin{equation}\label{eq:qkm-lyap}
  \HH^{(QKM)} = \frac{K}{2} \sum_{m,n} A_{m,n} \cos(\theta_m - \theta_n) +
  \frac{K_s}{2} \sum_m \cos^2(\theta_m).
\end{equation}
When the phase distribution is binary, $\theta_m = \sigma_m\pi/2$ with
$\sigma_m = \pm 1$, $\HH^{(QKM)} $ turns into an Ising Hamiltonian. Since
$\HH^{(QKM)} $ decreases with time, the arrival at a state with
clustered phases can be expected to deliver an approximation to the
ground state.

\begin{figure}[tb]
    \centering
    \includegraphics[width=3.5in]{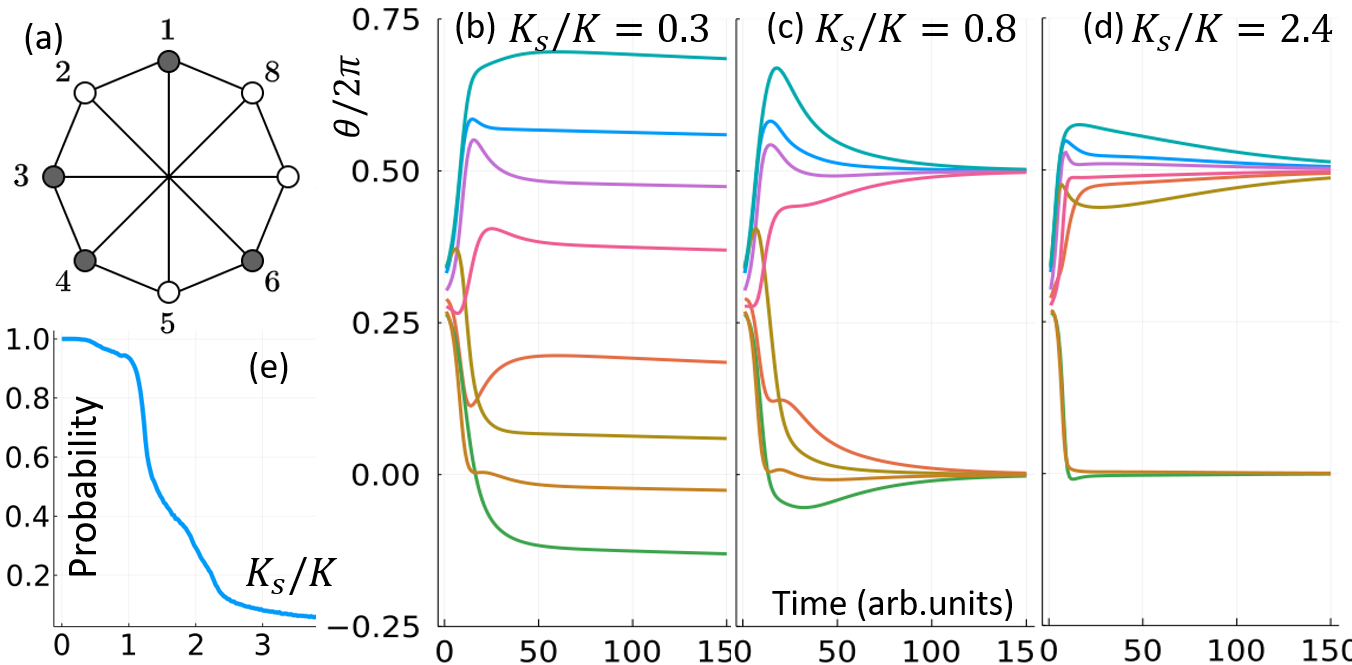}
    \caption{(a) A graph with $8 $ nodes and one of its maximal cut
      partitions. (b) - (d) Representative dynamical regimes obtained for
      different $K/K_s$. The obtained cut values are (c) 10, (d) 6. (e) The
      probability of finding a max-cut configuration as a function of
      $K/K_s $. }
    \label{fig:three-regimes}
\end{figure}

There are difficulties with such arguments: there is no guarantee that the
system will converge to a binary-like state, nor that the resultant state
will deliver the optimum solution. These difficulties manifest themselves
even on small graphs, like shown in Fig.~\ref{fig:three-regimes}(a) (the
same graph was considered in~\cite{mcquillan_oim_2019}). Depending on
$K_s/K $, the phase evolution governed by Eqs.~\eqref{eq:KM-full} can be in
any of three regimes:
\begin{enumerate}
\item well-defined clusters do not form;
\item clusters form and yield the maximal cut;
\item clusters form but do not yield the maximal cut. 
\end{enumerate}

It may appear as if QKM-based machines are unreliable
if $K_s$ does not ensure operating 
in regime 2. Moreover, as we will show, regimes 1 and 3 can be
regarded as generic.

However, this is regime 1, without well formed phase clusters, which
delivers the solution. It suffices that the network \emph{admits} the
binary state without necessarily settling
in it. Consequently, obtaining a steady state of a dynamic Ising
machine must be followed by rounding: finding the
best phase reference point. Forcing phases to form well-defined
clusters does not achieve this. This is supported by Fig.~\ref{fig:three-regimes}(e)
showing that when $K_s$ is too large,
the probability of success reduces to random guessingn.

\section{Ising model and the max-cut problem}

Let $\set{G} = \left\{ \set{V}, \set{E} \right\} $ be an undirected
graph on the sets of nodes $\set{V} $ and edges $\set{E}$. We will
denote the number of nodes and edges by $N = \abs{\set{V}} $ and
$M = \abs{\set{E}} $, respectively. The spin configurations are
described by binary functions on the graph:
$\bosi : \set{V} \to \left\{ -1,1 \right\}$. In other words, to each
node, we assign a binary variable $\sigma_m \in \left\{ -1,1 \right\}$.

With each configuration, an energy is associated
\begin{equation}\label{eq:I-ham}
  \quantop{H}(\bosi) = \sum_{(m,n) \in \set{E}} A_{m,n} \sigma_m \sigma_n
        = \frac{1}{2} \bosi \cdot \widehat{A} \cdot \bosi 
        = \frac{1}{2} \Tr[\widehat{A} \, \widehat{\Xi}],
\end{equation}
where $\widehat{A} $ is the graph adjacency matrix, and
$\widehat{\Xi} = \bosi \otimes \bosi $, so that $\Xi_{m,n} = \sigma_m \sigma_n $.
 In general, a weight function
$\widehat{J} : \set{E} \to \mathbb{R} $ can also be provided. We will focus on the
simpler case described above and only briefly discuss what changes in the
general case.

Each configuration $\bosi $ naturally defines partitioning
$\set{V} = \set{V}_+ \cup \set{V}_- $, with $\set{V}_+ $ and $\set{V}_- $
being sets, where $\bosi $ takes values $+1 $ and $-1 $, respectively.
Conversely, any partitioning, $\set{V} = \set{V}_+ \cup \set{V}_- $, defines a
configuration and, thereby, can be characterized by the energy, which is
related to the cut of $\set{G} $ induced by the partitioning. A cut,
$\set{C}(\set{V}_+, \set{V}_-)$ is a set of edges with one of the ends
in $\set{V}_+ $ and another in $\set{V}_- $. The cut size,
$C(\bosi) = \abs{\set{C}} $, can be found by summing over all edges a
function equal to $1 $ on edges from $\set{C} $, and to $0 $ elsewhere. In
terms of $\bosi $, the value of this function on edge $(m,n) $ can be
written as $(1 - \sigma_m \sigma_n)/2 $ leading to \cite{barahona_application_1988}
\begin{equation}\label{eq:cut-h}
 C(\bosi) = \frac{1}{2} \sum_{(m,n) \in \set{E}} A_{m,n}\left( 1 - \sigma_m \sigma_n \right) 
        = \frac{M}{2}  - \frac{1}{2} \HH(\bosi).
\end{equation}
Thus, finding the maximal cut,
$\bar{C}_{\set{G}} = \max_{\bosi \in {\left\{ -1,1 \right\}}^N}
C(\bosi)$ is equivalent to finding the ground state energy.

Owing to $A_{m,m} = 0$, finding the maximal cut amounts to finding a
maximal value of a linear function on the vertices of hypercube
${\left[ -1, 1\right]}^N$. In turn, any linear function on a convex
polyhedron reaches its extrema on the vertices. This
yields a continuous representation of the max-cut
problem~\cite{dezaGeometry1997}, when instead of binary functions
$\bosi$ one considers $\boxi : \set{V} \to [-1, 1]^N$:
\begin{equation} 
  \label{eq:max-cut-continuous-1} 
\bar{C}_{\set{G}} = \max_{\boxi \in {\left[ -1,1 \right]}^N} C(\boxi).
\end{equation}

Let us briefly consider a dynamical Ising machine utilizing this
representation. Away from the extreme points ($\xi_m = \sigma_m = \pm
1$), the machine's equations of
motion are
\begin{equation}\label{eq:cont-IM-eqm}
 \dot{\xi}_m = \frac{\partial C(\boxi)}{\partial \xi_m} = - \sum_n A_{m,n} \xi_n.
\end{equation}
Solutions obtained with the help of this machine are determined by the
machine's final state. It is easy to show that these are binary
configurations ($\xi_m = \sigma_m = \pm 1$) characterized by the majority
rule: for each node at least half of the incident edges are cut. Or,
equivalently, $F_m \geq 0$, where
\begin{equation}\label{eq:one-MR}
 F_m = - \sum_n A_{m,n} \sigma_m \sigma_n .
\end{equation}
We limit ourselves to the case of graphs without nodes with
even degrees (the number of incident edges). In this case, \emph{any} state
satisfying the majority rule is stable and has the basin of attraction
of finite volume. Thus, such Ising machine finds the same
configurations as a simple local search algorithm with the respective
consequences for performance.

In view of this consideration, it must be emphasized that the dynamical
Ising machines based on synchronized oscillator networks implement a
different operational principle and are not bound by the performance
limitations typical for the local search.

\section{QKM, XY model, and rank-2 SDP}

An extension of the Ising model, the XY model, is obtained by considering
vector-valued functions on the graph: $\boxi : \set{V} \to \mathbb{S}^2 $, where
$\mathbb{S}^2 $ is the set of unit vectors on a 2D Euclidean plane,
$\mathbb{R}^2 $. The configuration energy is given by
\begin{equation}\label{eq:xy-ham}
    \HH^{(XY)} (\boxi) = \frac{1}{2} \Tr[\widehat{A} \, \widehat{\Xi}^{(XY)}] 
                - \frac{K_s}{2} \sum_{m \in \set{V}} \vec{\xi}_m \cdot \tensor{l} \cdot \vec{\xi}_m ,
\end{equation}
where $\Xi^{(XY)}_{m,n} = \vec{\xi}_m \cdot \vec{\xi}_n$, $\tensor{l} = \vec{l} \otimes \vec{l}$
is the anisotropy tensor, $\vec{l}$ is the anisotropy axis, and $K_s $ is the
anisotropy constant. Depending on whether $K_s = 0 $ or $K_s \ne 0 $, the XY model
is called isotropic or anisotropic. Since $\vec{\xi}_m $ are 2D vectors (rather
than, say, 3D), the sign of $K_s $ plays no role. We assume that $K_s \geq 0$ so
that the anisotropy aligns the spins along the line defined by $\vec{l} $.

Treating $\HH^{(XY)}(\boxi) $ as a Lyapunov function, the dynamics is
defined by
$\dot{\vec{\xi}}_m = -\partial \HH^{(XY)}/\partial \vec{\xi}_m = g_m(\boxi) \vec{\gamma}_m $,
where $\vec{\gamma}_m $ is any of the two unit vectors orthogonal to
$\vec{\xi}_m $, and
\begin{equation}\label{eq:gm}
  g_m(\boxi) = - \sum_n A_{m,n} \vec{\gamma}_m \cdot \vec{\xi}_n +
  K_s \vec{\gamma}_m \cdot \tensor{l} \cdot \vec{\xi}_m.
\end{equation}
Equations of motion in this form ensure that $\abs{\vec{\xi}_m} $ is an
integral of motion. Alternatively, this can be enforced by
representing
$\vec{\xi}_m = \left( \cos(\theta) \, \sin(\theta) \right)^T$ in terms of the
angle $\theta_m$, say, with respect to $\vec{l} $. Using this
in Eq.~\eqref{eq:xy-ham}, we obtain
Eq.~\eqref{eq:qkm-lyap}.

Thus, QKM describes the dynamics of vector spins in the XY
model~\cite{shinomoto_phase_1986}. This relationship provides a
convenient phenomenology for discussing the dynamic
Ising machines implementing QKM as it abstracts from the challenges of
physical realizations of oscillator networks. Therefore,
for brevity, we will refer to these networks as XY machines.

The equivalence $\mathrm{QKM} \leftrightarrow \mathrm{XY}$ implies that the network
evolution realizes the gradient descent for the XY model. Generally,
the outcome of the evolution is a configuration with an unconstrained
mutual orientation of vector spins. This poses two questions:
\begin{enumerate}
\item  What is the relation between the final state of the XY model and the
ground state of the Ising model?
\item How to reconstruct a feasible
binary distribution from the ensemble of arbitrarily oriented spins?
\end{enumerate}

Finding the maximal cut (or the
ground state of the Ising model) can be formulated as an integer program:
\begin{equation}\label{eq:maxcut-ilp}
  \bar{C}_{\set{G}} = \frac{M}{2} -
      \min_{\widehat{\Xi}} \frac{1}{4} \Tr[\widehat{A}\, \widehat{\Xi}]
\end{equation}
with constraints $ \Xi_{m,m} = 1 $, and $\rank(\Xi) = 1 $. This problem is
APX-hard~\cite{papadimitriou_optimization_1991, khot_optimal_2004}
meaning that, unless $\textrm{P} = \textrm{NP}$, there is no
polynomial-time algorithm providing arbitrarily good approximation.
For such problems, the algorithms are characterized by the
approximation ratio:
$\rho = C_{\set{G}}(\boxi_g) /\bar{C}_{\set{G}} $, where
$C_{\set{G}}(\boxi_g)$ is the best solution the algorithm is
guaranteed to produce in polynomial time. For example, for local search
algorithms one has
$\rho \gtrsim 0.5$~\cite{korte_combinatorial_2018}. In particular, this holds
for the example of a dynamical machine driven by
Eq.~\eqref{eq:cont-IM-eqm}. Of course, due to the special form of the
respective worst-case Hamiltonians (see, for
instance,~\cite{kalantari_quadratic_1986}), this does not preclude
local search algorithms from performing very well on some classes of
graphs, which may be of practical relevance.

Other approaches to solving~\eqref{eq:maxcut-ilp} are based on
simplifying the problem by relaxing constraints and, subsequently,
reconstructing a feasible (satisfying the original constraints)
configuration by rounding. Importantly, since the relaxation cannot
change the problem complexity class, the complexity is delegated to
the rounding stage, which therefore requires special attention (see
e.g.~\cite{laurent_positive_1995, delorme_combinatorial_1993}).

It was discovered in~\cite{goemans_879-approximation_1994,
  goemans_improved_1995} that semidefinite programming (SDP) relaxation is
uniquely efficient in solving the max-cut problem. This relaxation requires
that $\widehat{\Xi}$ is symmetric positive semidefinite with
$\rank[\Xi] = k > 1$ (rank-$k$ relaxation). This is equivalent to
considering configurations $\boxi : \set{V} \to \mathbb{S}^k$ and minimizing
$\HH^{(k)}(\boxi) = \sum_{\set{E}} A_{m,n} \vec{\xi}_m \cdot \vec{\xi}_n $.
Given the solution of the relaxed problem, the feasible configuration is
obtained as $\sigma_m = \mathrm{sign}(\vec{\xi}^{(m)} \cdot \vec{t}) $, where
$\vec{t} \in \mathbb{S}^k $. Averaging the obtained cuts over randomly chosen
$\vec{t} $ results in~\cite{goemans_improved_1995,alon_bipartite_2000}
\begin{equation}\label{eq:gw-approx}
    \expval{C_{\set{G}}(\boxi)}_{\vec{t}} \geq \rho_{\mathrm{GW}} \bar{C}_{\set{G}},
\end{equation}
where $\rho_{\mathrm{GW}} =  \min_{\theta > 0} \theta/(\pi \sin^2(\theta/2)) \gtrsim 0.878 $.

The significance of this result is two-fold. First of all, it establishes a
rounding procedure recovering the solution in polynomial time. Second,
Eq.~\eqref{eq:gw-approx} estimates (not necessarily tightly) the so-called
integrality gap~\cite{raghavan_randomized_1987, gartner_approximation_2012}
and exhibits a relation between the solutions of the relaxed and integer
problems. This result holds for a weighted Ising model with weights of the
uniform sign. For models with variable signs, the standard argument fails,
but a guaranteed approximation ratio can still be
proven~\cite{charikar_maximizing_2004, alon_approximating_2004,
  anjos_strengthened_2002}.

To put this estimate into perspective, if $\mathrm{P} \ne \mathrm{NP}$, the
best classically possible approximation ratio for the max-cut problem is
$16/17 \gtrsim 0.941$~\cite{hastad_optimal_2001}. However, assuming additionally
the unique games conjecture~\cite{trevisanKhot2012, khotUnique2010}, it can
be shown that $\rho_{\mathrm{GW}}$ is the best approximation achievable in
polynomial time~\cite{khot_optimal_2004, raghavendraOptimal2008a,
  trevisanMaxCutSmallest2012, sotoImprovedAnalysisMaxCut2015}.

Importantly, we have the equivalence $\text{XY} \leftrightarrow \text{SDP}_{2}$, where
$\text{SDP}_2$ is the
rank-$2$ SDP relaxation~\cite{burer_rank-two_2002}. Indeed, for such
relaxation, one has
$\widehat{\Xi} = \vect{s}^{(1)} \otimes \vect{s}^{(1)} + \vect{s}^{(2)} \otimes
\vect{s}^{(2)}$ and hence
$\Xi_{m,n} = s_n^{(1)} s_m^{(1)} + s_n^{(2)} s_m^{(2)} = \vec{\xi}_n \cdot
\vec{\xi}_m$, where $\vec{\xi}_m = (s_m^{(1)}, s_m^{(2)})^T $ are unit
(because of the constraint $\Xi_{m,m} = 1 $) vectors.

This identifies the origin of the computational capabilities of XY
machines and, hence, Ising machines based on synchronizing oscillator
networks. Their dynamics implements the gradient descent minimization
of a rank-$2$ relaxation of the Ising ground state problem. Instead of
exploring the configuration space, as, say, is done in the Ising
machine described by Eq.~\eqref{eq:cont-IM-eqm}, the XY machines solve
a different but tightly related problem.

\begin{figure}[htbp]
\centering
\includegraphics[width=2.5 in]{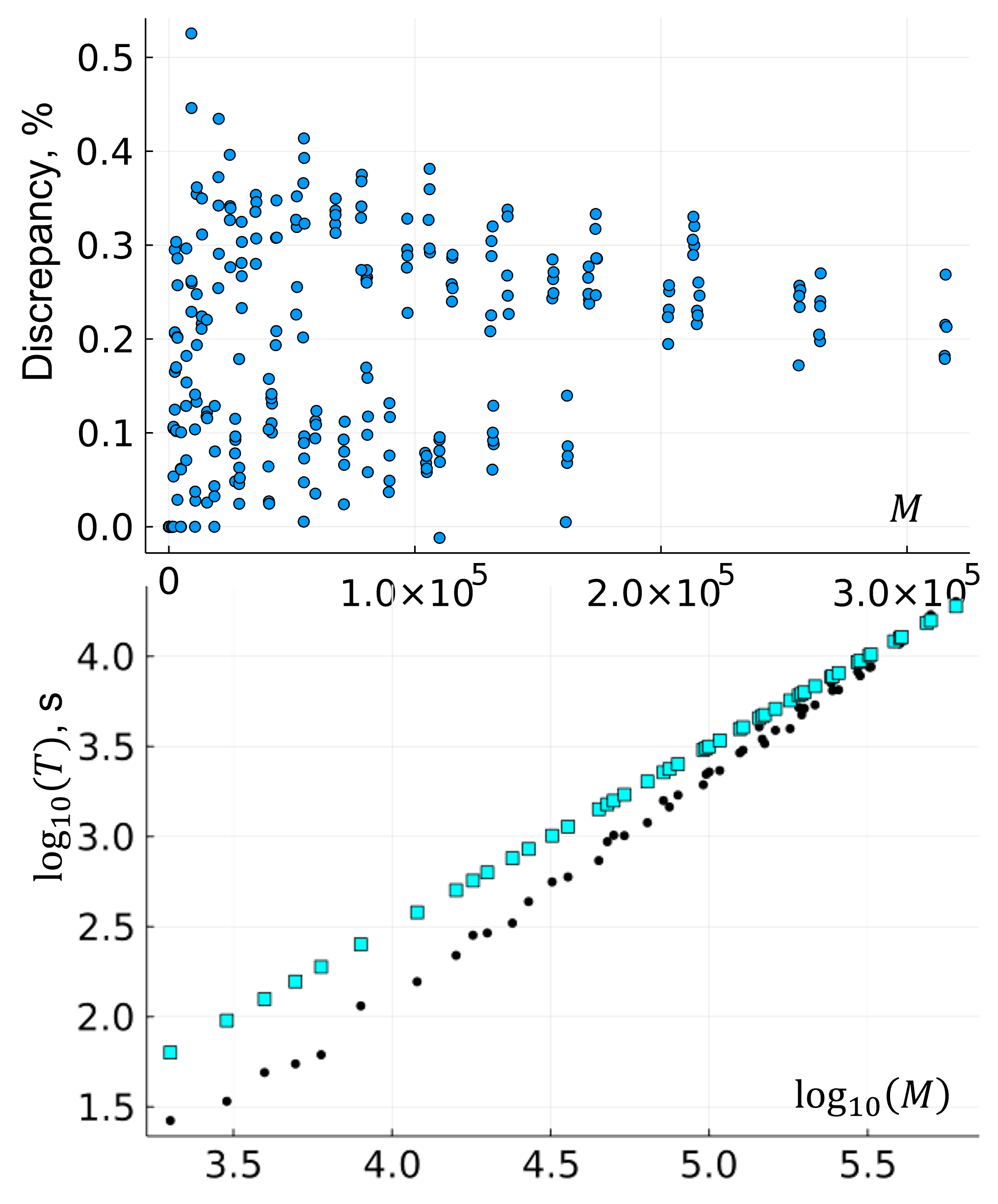}
\caption{Performance of a QKM-based dynamical Ising machine. (a)
  Discrepancy between the Ising machine and \texttt{Circut} results.
  (b) Scaling of the running wall-time with the number of edges in the
  graph (black circles) vs. the true linear scaling (cyan squares). }
{\label{fig:xy-scaling}}
\end{figure}

Figure~\ref{fig:xy-scaling} illustrates the performance of the
dynamical Ising machine implementing QKM. The dynamics of the machine
was taken to be governed by Eq.~\eqref{eq:KM-full} with $K_S = 0$ and
solved using the first order Euler approximation. The machine
ran on a series of Erd\H{o}s-R\'{e}nyi graphs $\set{G}_{N,p}$, where
$p$ is the probability for an edge to present, with
$50 \leq N \leq 2000$ and $0.1 \leq p \leq0.3$. For each graph, the machine ran
for $350$ time-steps each $20 K/N$ long. After each run, the final
configuration was rounded using the same algorithm as in
\texttt{Circut}~\cite{burer_rank-two_2002} and post-processed as
described below. This procedure was repeated $300$ times from
independently chosen random configurations, and the best value of cut
was recorded. The obtained cut values were compared with
\texttt{Circut} results, which remains one of the best max-cut
heuristic solvers~\cite{dunningWhat2018}. Except for a single
instance, the results obtained by the Ising machine were within 0.5
percent of \texttt{Circut}'s.

As an estimate of running time, the wall-time, $T$, was measured
(Fig.~\ref{fig:xy-scaling}). It must be noted that, by design, the
number of elementary operations in simulating the machine dynamics is
$\mathcal{O}(M)$. The observed deviation from this scaling is due to the
rounding procedure, which, as implemented, scales at worst as
$\mathcal{O}(NM)$.

The post-rounding processing consisted of two steps based on the observation
that rounding does not necessarily respect the majority rule. The
first step implemented the local search ensuring that all nodes obey
$F_m \geq 0$. The second step, relevant for nodes with $F_m = 0$, ensured
that for each cut edge at least half adjacent edges should be cut
(otherwise, the cut can be increased by reverting spins at the
incident nodes).

\section{The detrimental effect of forced binarization}

It must be emphasized that the chain of equivalences
$$\text{QKM} \leftrightarrow \text{XY} \leftrightarrow \text{SDP}_2$$
concerns
only the dynamics of the Ising machines. The principal part of
accessing the computational resource associated with SDP is to
properly recover a feasible binary state from the configuration of
unit vectors of the XY model or relative oscillator
phases in a synchronized network. Rounding the state
of the Ising machines remains an under-explored problem since the existing
implementations pursue dynamics collapsing the machine to a binary
state. This can be achieved when the anisotropy is sufficiently
strong. At the same time, as demonstrated by
Fig.~\ref{fig:three-regimes}e, strong anisotropy may disrupt finding
the configuration delivering the maximal cut. Here, we show that such
detrimental effect of strong anisotropy is generic.

The equilibrium configurations of the oscillator networks implementing
QKM are determined by $g_m(\boxi^{(0)}) = 0 $. The dynamics of weak
excitations is obtained by representing
$\vec{\xi}_m = \vec{\xi}_m^{(0)} + x_m \vec{\gamma}_m^{(0)} $, with
$x_m \ll 1 $, so that
\begin{equation}\label{eq:xy-weak}
\begin{split}
\dot{x}_m =
    & \sum_n A_{m,n} \vec{\xi}_m^{(0)} \cdot \vec{\xi}_m^{(0)} (x_m - x_n) \\
        & - K_s \left(2 \vec{\xi}_m^{(0)} \cdot \tensor{l} \cdot\vec{\xi}_m^{(0)} - 1 \right) x_n,
\end{split}
\end{equation}
or
$\dot{\vect{x}} = \left[ \widehat{L}(\boxi^{(0)}) -
  \widehat{K}(\boxi^{(0)}) \right] \vect{x} $.

The dynamics is attracted to (Lyapunov) stable equilibria, that is
with \emph{negative} semidefinite
$\widehat{L}(\boxi^{(0)}) - \widehat{K}(\boxi^{(0)})$. We note that
the Laplacian structure of $\widehat{L}(\boxi^{(0)}) $ implies that it
always has a zero eigenvalue. As a result, in the isotropic case, weak
perturbations of stable configurations exponentially converge to their
projection on the homogeneous displacement, $x_m \equiv x $, which reflects
the rotational symmetry of the XY model.

It must be emphasized, that because of the nonlinear coupling between
oscillators in QKM, matrix $\widehat{L}(\boxi^{(0)})$ coincides with
the graph Laplacian \emph{only} when all $\vec{\xi}^{(0)}$ have the same
orientation. For other configurations, the spectral properties of the graph
Laplacian and $\widehat{L}(\boxi^{(0)})$ are drastically different.

It follows from~\eqref{eq:gm}, that $g_m(\boxi) $ vanishes on
Ising-like configurations $\boxi^{(I)} $ with
$\vec{\xi}_m^{(I)} = \sigma_m \vec{l}$. It is straightforward to show that in
the isotropic case, all configurations $\bosi$ that do not produce the
maximal cut are unstable~\cite{burer_rank-two_2002}. In turn, maximal
cut configurations are stable only on bipartite graphs and selected
families of non-bipartite graphs. As a result, anisotropy plays
the major role in the emergence of Ising-like configurations in the
dynamics of XY machines.

A complete framework describing the effect of anisotropy on the
convergence properties of XY machines is yet to be developed. In the
present paper, we limit ourselves to an analysis of the structure of
the binary configurations enabled by anisotropy.

Since the effect of anisotropy on binary configurations reduces to a
simple displacing the spectrum of $\widehat{L}(\bosi) $ by $-K_s $,
any such configuration can be stable provided
\begin{equation}\label{eq:Ks-stability}
 K_s \geq \kappa(\bosi) = \lambda_1(\widehat{L}(\bosi)),
\end{equation}
where we have introduced $\kappa(\bosi)$, the \emph{instability} of
$\bosi$, which is defined by $\lambda_1(\widehat{L}(\bosi))$, the maximal
eigenvalue of $\widehat{L}(\bosi) $. Thus, except for
special graphs, the instability of binary configurations of the XY
model is positive.

Based on this, we can identify characteristic values of anisotropy,
when the significant impact on the computational capabilities can be
expected. The condition
$\widehat{L}(\bosi) - \widehat{K}(\bosi) \preccurlyeq 0$ implies that
for any unit $\mathbf{u} \in \mathbb{R}^N$, one has
$K_s - \mathbf{u} \cdot \widehat{L}(\bosi) \cdot \mathbf{u} \geq 0$, or that
\begin{equation}\label{eq:}
K_s + \sum_m u_m^2 F_m + \sum_{m,n} A_{m,n} u_m u_n \geq 0,
\end{equation}
where $F_m$ are given by Eq.~\eqref{eq:one-MR}.

This yields the first characteristic value of anisotropy,
$K_s^{(1)} = \abs{\mu_N(\widehat{A})}$, where $\mu_N(\widehat{A})$ is the
smallest eigenvalue of the graph adjacency matrix. When $K_s$ reaches
$K_s^{(1)}$, any binary states satisfying the majority rule
$F_m \geq0$ become stable. Using a simple bound
$\abs{\mu_N(\widehat{A})} \leq \Delta_{\set{G}}$, where
$\Delta_{\set{G}}$ is the graph maximal degree, the following rule can be
formulated. When $K_s = \Delta_{\set{G}}$, the structure of binary
configurations produced by the anisotropic XY machine is, at best, the
same as obtained by simple local search.

Clearly, when anisotropy increases further, binary configurations that
do not satisfy the majority rule (and, hence, cannot maximize cut)
also belong to the set of stable configurations. In other words, the
quality of the solutions may become worse than that of the local
search. Finally, when anisotropy is too strong,
$K_s \geq K_s^{(2)} = \lambda_1(\widehat{L}) $, where
$\lambda_1(\widehat{L}) $ is the maximal eigenvalue of the graph Laplacian,
\emph{all} binary configurations are stable. For instance, this is the
case when $K_s \geq 2 \Delta_{\set{G}}$.

While identifying ``dangerous'' values of the anisotropy, this
consideration leaves open the question whether moderate anisotropy
$ K_s < \Delta_{\set{G}}$ can be used to arrive at a maximal-cut binary
configuration.

\begin{figure}[tb]
    \centering
    \includegraphics[width=3.5in]{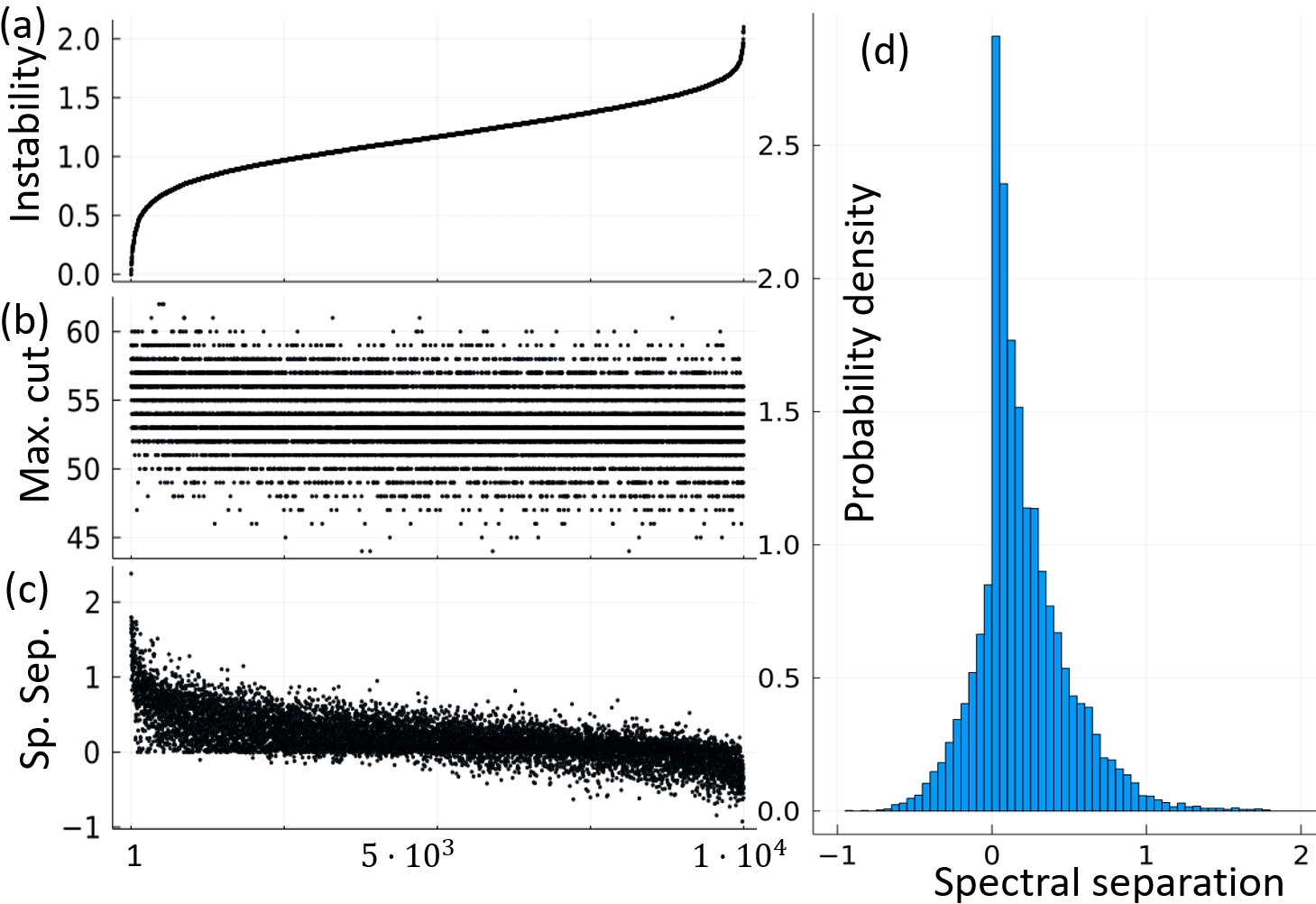}
    \caption{(a) The instability of the ground (max-cut) state, (b) the maximal cut, and
      (c) the spectral separation $\delta_{\set{G}}$ for $10^4 $ connected Erd\H{o}s-R\'{e}nyi graphs
      $\set{G}_{17, 0.7} $. The samples are arranged according to the
      instability. (d) The empirical probability distribution function of $\delta_{\set{G}}$.}
    \label{fig:sp-gap}
\end{figure}

A parameter characterizing how widely one can vary $K_s $ without
introducing sub-optimal configurations is the difference between
instabilities of the least unstable maximal cut and non-max-cut
configurations
\begin{equation}\label{eq:sp_sep}
    \delta_{\set{G}} = \min_{\bosi \in \set{M}_{\set{G}}} \kappa(\bosi)
           - \min_{\boxi \notin \set{M}_{\set{G}}} \kappa(\bosi),
\end{equation}
where $\set{M}_{\set{G}}$ is the set of max-cut configurations.
Importantly, neither the magnitude of $\delta_{\set{G}} $ nor even its sign
are bounded on a sufficiently rich set of graphs. As an example of
such set we have considered $10^4 $ connected random (Erd\H{o}s-R\'{e}nyi)
graphs $\set{G}_{17, 0.7}$. Figure~\ref{fig:sp-gap} shows the
instability of the max-cut configuration together with the max-cut
values (Fig.~\ref{fig:sp-gap}b) and spectral separations
$\delta_{\set{G}}$ (Fig.~\ref{fig:sp-gap}c). It reveals that there is only
a weak correlation between the instabilities of the max-cut and
non-max-cut configurations. This is summarized in
Fig.~\ref{fig:sp-gap}d depicting the distribution function of
$\delta_{\set{G}}$. It shows that the probability that in a randomly chosen
graph the max-cut-state will become stable at the lowest value of
anisotropy is rather moderate ($0.2$ for the considered set).

These simulation results show that the probability to have a graph admitting
anisotropy governed selection of the ground state is small, which
supports our statement that regimes $1$ and $3$ in
Fig.~\ref{fig:three-regimes} should be regarded as generic.

\section{Conclusion}

We have shown that the computational resource of dynamic Ising
machines based on synchronizing networks of nonlinear oscillators
originates from the factual realization of rank-$2$ semidefinite
programming relaxation of the max-cut problem. In contrast to
approaches aiming at direct exploration of the Ising model space
state, these relaxations deliver the best (if
$\mathrm{P} \ne \mathrm{NP}$) approximation achievable in polynomial
time. This shows that Ising machines based on synchronizing networks
are capable of providing good heuristics for a wide class of NP-hard
problems.

At the same time, this relation shows that to reach theoretically
possible performance, a rounding procedure must be supplied. An attempt to
force the system to evolve towards an Ising-like state may disrupt the
computational capabilities up to their complete invalidation when the dynamic
Ising machine effectively acts as a random generator of
configurations. For the Ising machines utilizing an effective
anisotropy for ensuring final binary states, we have estimated the
critical values of anisotropy corresponding to the loss of quality of
solutions obtained by the collapsed state.

Finally, our consideration demonstrates that a quantitative evaluation of
dynamic Ising machines requires an accurate description of their dynamics. The
ability to yield the Ising Hamiltonian is not enough because of the wide
variability of the approximation ratios: from the best classically possible to
that of a random generator.

\section*{Acknowledgements}

The work was supported by the US National Science Foundation under Grant No.
1710940 and by the Air Force Office of Scientific Research (AFOSR)
under Grant No. FA9550-16-1-0363.


\end{document}